%% file: WWW2019-DLoB.tex
\documentclass[sigconf]{acmart}

\usepackage{booktabs} 
\usepackage{multirow,color, hyperref, url, graphics}
\usepackage{subfigure}
\usepackage{listings}
\usepackage{xcolor}
\usepackage{amsmath}
\usepackage{diagbox}

\setcopyright{none}
\copyrightyear{2019}
\acmYear{2019}
\acmConference[WWW '19]{Proceedings of the 2019 World Wide Web Conference}{May 13--17, 2019}{San Francisco, CA, USA}
\acmBooktitle{Proceedings of the 2019 World Wide Web Conference (WWW'19), May 13--17, 2019, San Francisco, CA, USA}
\acmPrice{}
\acmDOI{10.1145/3308558.3313639}
\acmISBN{978-1-4503-6674-8/19/05}

\usepackage{balance}

\begin{document}

\title{Moving Deep Learning into Web Browser: How Far Can We Go?}

\author{Yun Ma$^{1,2}$, Dongwei Xiang$^1$, Shuyu Zheng$^1$, Deyu Tian$^1$, Xuanzhe Liu$^1$}
\affiliation{%
  \institution{{\small{$^1$}}Key Lab of High-Confidence Software Technology, MoE (Peking University), Beijing, China
  \\{\small{$^2$}}Tsinghua University, Beijing, China}
}
\email{{mayun, xdw, zhengshuyu, tiandeyu, xzl}@pku.edu.cn}
%
%
%
%
%

\begin{abstract}
Recently, several JavaScript-based deep learning frameworks have emerged, making it possible to perform deep learning tasks directly in browsers. However, little is known on what and how well we can do with these frameworks for deep learning in browsers. To bridge the knowledge gap, in this paper, we conduct the first empirical study of deep learning in browsers. We survey 7 most popular JavaScript-based deep learning frameworks, investigating to what extent deep learning tasks have been supported in browsers so far. Then we measure the performance of different frameworks when running different deep learning tasks. Finally, we dig out the performance gap between deep learning in browsers and on native platforms by comparing the performance of TensorFlow.js and TensorFlow in Python. Our findings could help application developers, deep-learning framework vendors and browser vendors to improve the efficiency of deep learning in browsers.
\end{abstract}

%
%


\maketitle

\section{Introduction}\label{sec:introduction}
\input{sections/introduction}

\section{Background}\label{sec:background}
\input{sections/background}

\section{Supported Features of Deep Learning in Browsers}\label{sec:methodology}
\input{sections/support.tex}

\section{Performance of Deep Learning in Browsers}\label{sec:per}
\input{sections/performance.tex}

\section{Comparison with Native Framework}\label{sec:compare}
\input{sections/comparison.tex}

\section{Implications}\label{sec:implications}
\input{sections/implication}

\section{Related Work}\label{sec:related}
\input{sections/related}

\section{Conclusion}\label{sec:conclusion}
\input{sections/conclusion}

\begin{acks}
This work was supported by the National Key R\&D Program of China under the grant number 2018YFB1004800, the National Natural Science Foundation of China under the grant number 61725201, the Beijing Municipal Science and Technology Project under the grant number Z171100005117002, and China Postdoctoral Science Foundation.
\end{acks}

\bibliographystyle{ACM-Reference-Format}
\balance
\bibliography{dlob}

\end{document}

%% file: sections/introduction.tex
In the past decade, the advance of deep learning (DL) technique has significantly promoted the artificial intelligence (AI). Numerous AI applications, e.g.,  image processing, object tracking, speech recognition, and natural language processing, have raised urgent requirements to adopt the DL. As a result, various libraries and frameworks, such as TensorFlow~\cite{abadi2016tensorflow}, Caffe~\cite{Caffe}, and CNTK~~\cite{CNTK}, have been proposed and applied in practice.

However, developing AI applications powered by the popular DL frameworks and libraries is a non-trivial task. Usually, these frameworks and libraries are leveraged by \textit{native} applications that can run on heterogeneous development environments such as Windows, Linux, MacOS/iOS, and Android. The applications are developed by various imperative programming languages, i.e.,  C/C++ on Windows, Objective-C on iOS and MacOS, and Java on Android. Developing AI applications that is portable to multiple platforms is indeed not easy. The development is particularly complicated for mobile applications, as the app vendors usually need to develop and maintain both iOS and Android versions. In addition, the deployment is also non-trivial, as most current platforms come with an appstore, some of which require manual testing of submitted applications by the appstore provider before being published-a process that can take several weeks-and applications can be rejected for seemingly arbitrary reasons.

Compared to the native applications, Web applications can indeed make the cross-platform portability issues much simpler. The same implementation of a DL-powered Web application can be deployed in the browser on all platforms regardless of the underlying hardware device types (PC, smartphones, and wearable devices) and the operating systems (Windows, Mac, iOS, and Android). Advancements in HTML5, CSS3, and especially JavaScript language, started to enable the creation of DL-powered Web applications that offer a comparable experience to native applications, especially for the popular Web game applications~\cite{tucker2018inverse}\cite{Oh2015ActionConditionalVP}. In particular, benefited from the development of WebGL~\cite{marrin2011webgl}\cite{auer2012real}\cite{chen2011framework}, current major browsers such as Google Chrome, Mozilla FireFox, and Apple Safari, can better utilize the integrated graphics card to accelerate DL tasks, without the need of standalone graphics card like NVIDIA which is required by native DL frameworks.

Running DL-powered Web applications in browsers has drawn the attention from various research communities including AI, software engineering, Web browsers, and even computer architecture. As a result, various JavaScript-based DL development frameworks and libraries have been published. In 2015, Karpathy presented the ConvNetJS~\cite{ConvNetJS}, known as the first JavaScript library for DL in Web browsers to date. Other efforts such as
WebDNN~\cite{WebDNN}, Keras.js~\cite{Keras.js}, and Mind~\cite{Mind}, were proposed to support DL in browsers. In early 2018, Google released the TensorFlow.js~\cite{TensorFlow.js}, which is a significant step for promoting the DL in browsers.




Although the preceding efforts along with some on-going efforts seem to make running DL tasks in browsers possible, little is known on what DL tasks we can do and how well DL works in browsers. More importantly, considering the long debate of performance of Web applications compared with that of native applications, the same issue also exists in developing DL-powered Web applications. Hence, it is urgent to address such a knowledge gap in terms of the feasibility and usability for running DL in Web browsers.

In this paper, we make the first empirical study of DL in browsers by answering the following research questions.
\begin{itemize}
    \item \textbf{RQ1}: What features do existing frameworks provide to implement various kinds of DL tasks in the browser?
    \item \textbf{RQ2}: How well do existing frameworks perform over different DL tasks?
    \item \textbf{RQ3}: How big is the performance gap between running DL in the browser and on the native platform?
\end{itemize}

We select 7 popular JavaScript-based frameworks that support running DL in browsers, and conduct a characteristic study over them. We develop a browser extension to measure the performance as well as the utilization of system resources when running different DL tasks. We choose the TensorFlow.js and native TensorFlow in Python to compare the performance of DL in browsers with that on native platforms.

The key findings of our study includes:

  \noindent $\bullet$ \textbf{DL in browsers is still at dawn.} Most frameworks of DL in browsers support only a specific subset of DL tasks. Among all the frameworks, TensorFlow.js provides the most number of functionalities to realize various kinds of DL tasks.

  \noindent $\bullet$ \textbf{Support of training in browsers is not fledged.} In most frameworks, inference has drawn more attention compared with training. For training tasks, the number of neurons per layer dominates the performance variation considering the complexity of DL models since the browser is limited in complex matrix calculation.

  \noindent $\bullet$ \textbf{Model loading dominates the computation for inference tasks.} Loading and warming up the DL model costs more time than running the inference task itself. The CPU backend performs better than the GPU backend when the browser run inference tasks for small-size models.

  \noindent $\bullet$ \textbf{Integrated graphics card helps browsers to beat native platforms when standalone graphics card is not available.} For popular pre-trained models like MobileNet and Inception, TensorFlow.js is just 1x to 2x slower than native TensorFlow in Python when running inference tasks on the standalone graphics card. TensorFlow.js on the integrated graphics card outperforms the native TensorFlow on CPU when running the same inference task.

  \noindent $\bullet$ \textbf{System resources can be further exploited for in-browser DL tasks.} For TensorFlow.js, the CPU is not fully utilized (about 80\%) when DL tasks run on the CPU backend. The memory allocated to WebGL is limited by the browser, leading to the crash of some DL tasks.

Based on the findings, We have drawn some practical recommendations for application developers as well as DL-framework and browser vendors. Application developers who aim to develop DL-powered Web applications, need to better control the the number of neurons per layer in DL models, pre-load the model file in advance, and employ the CPU backend rather than the GPU backend when running inference tasks on small DL models. DL-framework vendors should consider encoding the model file in binary format rather than JSON to reduce the file size as well as improve the computation time, and leverage compiler optimization techniques to reduce the call stack. Browser vendors should consider supporting multi-process and scheduling over multi-core in the JavaScript engines.

The remainder of this paper is organized as follows. Section~\ref{sec:background} shows some background knowledge of deep learning in browsers. Sections~\ref{sec:methodology} to \ref{sec:compare} describe the results, including the analysis of framework functionality, performance measurement, and comparison with native DL frameworks. Section~\ref{sec:implications} presents the implications and recommendations drawn from the findings. Section~\ref{sec:related} surveys related work and Section~\ref{sec:conclusion} concludes the paper with future work.

%% file: sections/background.tex
In this section, we give some background of deep learning and then discuss how browsers support deep learning tasks.

\subsection{Deep Learning}
Deep learning (DL) is a class of machine learning algorithms that use a cascade of multiple layers of nonlinear processing units (called neurons) for feature extraction and transformation. Each successive layer uses the output from the preceding layer as input. In recent years, DL has gained great success in many areas such as computer vision, speech recognition and natural language processing.

There are many types of DL models, among which deep neural network (DNN)~\cite{hinton2006fast}, convolutional neural network (CNN)~\cite{lecun1989generalization}, and recurrent neural network (RNN)~\cite{rumelhart1986learning} are three basic structures. DNN is a typically feedforward network with multiple layers between the input and output layers, in which data flows from the input layer to the output layer without looping back. CNN uses a variation of multi-layer perceptrons designed to require minimal preprocessing, usually applied to analyzing visual imagery. RNN has connections between nodes forming a directed graph along a temporal sequence, allowing it to exhibit temporal behaviors.

DL consists of two phases: training phase where the input data are used to calculate the parameters of the model, and inference phase where the trained model outputs the value given a specific input sample.

\subsection{Deep Learning in Browsers}

%
%

Recently, there is a trend that applications perform DL tasks directly on the clients for better privacy and timely response. As a cross-platform client-side computation target, Web browsers have drawn the attention to AI communities to support client-side DL. Several applications of in-browser DL are implemented and published, such as 1) TensorFlow playground~\cite{playground}, which is an interactive platform to learn the principle of DL; 2) Teachable Machine~\cite{teachablemachine}, which gives the users an experience of teaching the machine how to response when they pose a gesture, using camera in the browser; 3) MLitB ~\cite{10.7717/peerj-cs.11}, which is capable of performing distributed learning with heterogeneous classes of devices using Web browsers; 4) MorphCast~\cite{MorphCast}, which combines interactive video and face recognition with emotion, gender and age analysis to create adaptive-media.

DL in browsers is implemented by JavaScript and rely on the browser engine to execute. Fortunately, the advancement of latest browsers provides APIs to access GPU, which can be used to accelerate matrix calculations of DL. These APIs are: 1) \textbf{WebGL}~\cite{webgl}, which is a JavaScript API for rendering interactive 2D and 3D graphics within any compatible Web browser; 2) \textbf{WebGPU}~\cite{webgpu}, which is the fastest among existing JavaScript APIs for accelerating graphics and computation. Currently, WebGPU API is supported only in Safari Technology Preview. We should mention that WebGL and WebGPU can run on both integrated graphics cards and standalone graphics cards.




%% file: sections/support.tex
In this section, we make a characteristic study to answer the first research question, i.e., what features do existing frameworks provide to implement various kinds of DL tasks in the browser? We first introduce the frameworks selected for the study. Then we compare the features of these frameworks from two aspects: provided functionality and developer support. For provided functionality, we mainly examine whether each framework supports some basic functionalities that are commonly used in the development of DL applications. For developer support, we take a look at some factors which may affect the efficiency of developing and deploying DL applications. Table~\ref{table:frameworks} summarizes all the results as of Nov. 2018.

\begin{table*}
\centering
\caption{Characteristics of JavaScript-based frameworks that support deep learning in browsers.}\label{table:frameworks}
\vspace{-0.5em}
\small
\scalebox{0.94}{
\begin{tabular}{lcccccccc}
\hline
\multicolumn{2}{l}{} & \textbf{TensorFlow.js} & \textbf{ConvNetJS} & \textbf{Keras.js} & \textbf{WebDNN} & \textbf{brain.js} & \textbf{synaptic} & \textbf{Mind} \\
\hline
\hline
\multicolumn{9}{c}{\textbf{Basic Information}} \\
\hline
\hline
\textbf{Github Stars} &  & 9453 & 9364 & 4348 & 1464 & 6366 & 6315 & 1333 \\
\hline
\textbf{Main Contributor} &  & Google & \begin{tabular}[c]{@{}l@{}}Stanford \\ University\end{tabular} & Leon Chen & \begin{tabular}[c]{@{}l@{}}The University \\ of Tokyo\end{tabular}& \begin{tabular}[c]{@{}l@{}}Robert \\ Plummer\end{tabular} & \begin{tabular}[c]{@{}l@{}}Juan \\ Cazala\end{tabular} & \begin{tabular}[c]{@{}l@{}}Steven \\ Miller\end{tabular} \\
\hline
\textbf{Last Commit Date} &  & Oct 30, 2018 & Nov 25, 2016 & Aug 17, 2018 & Oct 25, 2018 & Nov 5, 2018 & Mar 25, 2018 & Jul 7, 2017 \\
\hline
\textbf{Status} &  & Active & Not Active & Not Active & Active & Active & Active & Not Active \\
\hline
\hline
\multicolumn{9}{c}{\textbf{Functionality}} \\
\hline
\hline
\textbf{Support for Training} &  & Y & Y & N & N & Y & Y & Y \\
\hline
\multirow{3}{*}{\begin{tabular}[c]{@{}l@{}}\textbf{Supported} \\ \textbf{Network Types}\end{tabular}}  & \textbf{DNN} & Y & Y & Y & Y & Y & Y & Y \\
\cline{2-9}
    & \textbf{CNN} & Y & Y & Y & Y & N & N & N \\
\cline{2-9}
    & \textbf{RNN} & Y & N & Y & Y & Y & Y & N \\
\hline
\multicolumn{2}{l}{\textbf{Supported Layer Types}}  & 49 & 7 & NA & NA & 7 & 1 & 1 \\
\hline
\multicolumn{2}{l}{\textbf{Supported Activation Types}}  & 16 & 4 & NA & NA & 4 & 5 & 2 \\
\hline
\multicolumn{2}{l}{\textbf{Supported Optimizer Types}} & 7 & 3 & NA & NA & 1 & NA & NA \\
\hline
\multicolumn{2}{l}{\textbf{Support for GPU Accelaration (WebGL)}} & Y & N & Y & Y & N & N & N \\
\hline
\hline
\multicolumn{9}{c}{\textbf{Developer Support}} \\
\hline
\hline
\textbf{Documents} &  & Y & Y & Not finished & Y & Only tutorials & Y & Y \\
\hline
\textbf{Demos} &  & 20 & 10 & 9 & 8 & 7 & 7 & 4 \\
\hline
\multirow{3}{*}{\begin{tabular}[c]{@{}l@{}} \textbf{Importing Models from} \\\textbf{Other Frameworks} \end{tabular}} & \textbf{TensorFlow} & Y & N & N & Y & N & N & N \\
\cline{2-9}
    & \textbf{Keras} & Y & N & Y & Y & N & N & N \\
\cline{2-9}
    & \textbf{Caffe\&Pytorch} & N & N & N & Y & N & N & N \\
\hline
\multirow{2}{*}{\textbf{API to Save/Load Model}} & \textbf{Save} & Y & Y & N & N & Y & Y & Y \\
\cline{2-9}
    & \textbf{Load} & Y & Y & Y & Y & Y & Y & Y \\
\hline
\multicolumn{2}{l}{\textbf{Support for Server Side (Node.js)}}  & Y & Y & Y & Y & Y & Y & Y \\
\hline
\textbf{Library Size} &  & 732KB & 33KB & 650KB & 130KB & 819KB & 106KB & NA \\
\hline
\end{tabular}
}
\end{table*}

\subsection{Selected Frameworks}
To select the state-of-the-art frameworks of supporting DL in browsers, we search on the GitHub with the keyword ``deep learning framework'' and filter the results in JavaScript language. Then we choose the top 7 frameworks of which the number of stars exceeds 1,000 on GitHub. We introduce each framework as follows.

\noindent \textbf{TensorFlow.js}~\cite{TensorFlow.js}, released by Google in Mar. 2018, is an in-browser machine learning library that supports defining, training, and running models entirely in the browser using JavaScript. It is the successor to deeplearn.js which is now called TensorFlow.js Core. TensorFlow.js is powered by WebGL and provides high-level APIs for defining models. TensorFlow.js support all the Keras layers (including Dense, CNN, LSTM, and so on). Therefore, it is easy to import models pre-trained by the native TensorFlow and Keras into the browser and run with Tensorflow.js.

\noindent \textbf{ConvNetJS}~\cite{ConvNetJS} is a Javascript library originally written by Andrej Karpathy at Stanford. The entire library is based on transforming 3-dimensional volumes of numbers. ConvNetJS currently supports common neural network models and cost functions for classification and regression. Furthermore, it supports convolutional networks, and an experimental reinforcement learning. Unfortunately, although ConvNetJS might be the most famous framework before TensorFlow.js, it is no longer maintained after Nov. 2016.

\noindent \textbf{Keras.js}~\cite{Keras.js} abstracts away a number of frameworks as backends including TensorFlow, CNTK, etc. It supports importing models pre-trained by Keras for inference. In the GPU mode, computation is performed by WebGL. However, this project is no longer active.

\noindent \textbf{WebDNN}~\cite{WebDNN}, released by the University of Tokyo, claims to be the fastest DNN execution framework in browsers. It supports only the inference tasks. The framework supports 4 execution backends: WebGPU, WebGL, WebAssembly, and fallback pure JavaScript implementation. WebDNN optimizes DNN models by compressing the model data to accelerate the execution. Empirical evaluations showed that it achieved more than 200x acceleration~\cite{webdnnbench}.

\noindent \textbf{brain.js}~\cite{brain.js} is a JavaScript library for neural networks replacing the deprecated ``brain'' library. It provides DNN, RNN, LSTM and GRU for training tasks. The library supports serializing and loading the state of a trained DL model with JSON.

\noindent \textbf{synaptic}~\cite{synaptic.js} is a JavaScript architecture-free neural network library, supporting basically any type of first order or even second order RNN. This library also includes a few built-in DL architectures, including multi-layer perceptrons, LSTM, liquid state machines and Hopfield networks.

\noindent \textbf{Mind}~\cite{Mind} is a flexible neural network library. The core framework has only 247 lines of code, which uses a matrix implementation to process training data. It supports customization of the network topology and plugins to configure pre-trained models created by the mind community. However, this framework is no longer active.

\subsection{Provided Functionality}

\noindent \textbf{Support for training.} Most frameworks support training and inference tasks in the browser. However, Keras.js and WebDNN do not support training DL models in browsers. They support only loading pre-trained models to perform inference tasks. Therefore, the number is not available for the types of layer/activation/optimizer supported by Keras.js and WebDNN in Table~\ref{table:frameworks}.

\noindent \textbf{Supported network types.} Some frameworks are not for general-purpose DL tasks, so they differ in the supported network types. Specifically, TensorFlow.js, Keras.js and WebDNN support three network types: DNN, CNN and RNN. However, ConvNetJS mainly supports CNN tasks and does not support RNN. brain.js and synaptic mainly support RNN tasks, and do not support convolution and pooling operations used in CNN networks. Mind supports only the basic DNN.

\noindent \textbf{Supported layer types.} All frameworks support building neural networks using units of layers. The layer API of TensorFlow.js supports 49 different layers, including dense, convolution, pooling, RNN, normalization, and so on. Other frameworks support a smaller variety of layers, which are also related to the network types they support. It should be noted that the core API of TensorFlow.js is implemented in a way similar to the native TensorFlow which combines various operations to build computational graphs. synaptic is an architecture-free framework that supports building any type of first order or even second order RNN networks.

\noindent \textbf{Supported activation/optimizer types.} In general, TensorFlow.js provides developers with the most kinds of choices. For activation functions, other frameworks support only basic sigmoid or ReLU. For optimizers, other frameworks mainly support basic stochastic gradient descent (SGD).

\noindent \textbf{Support for GPU acceleration (WebGL).} TensorFlow.js is the only framework that supports GPU-accelerated training tasks. TensorFlow.js, Keras.js, and WebDNN support using GPU to accelerate inference tasks. WebDNN also supports a more advanced technology, WebGPU, but WebGPU has been supported by only the technology preview version of Safari.

\subsection{Developer Support}

\noindent \textbf{Documents.} Documents provided by TensorFlow.js, ConvNetJS, WebDNN and synaptic are completed and in detail. The document of Keras.js is not complete and brain.js has only a few tutorials.

\noindent \textbf{Demos.} All the frameworks provide demos for developers to get start. TensorFlow.js offers the richest demos covering a wide range of use cases.

\noindent \textbf{Importing models from other frameworks.} TensorFlow.js, Keras.js and WebDNN support importing models from native DL frameworks in Python and all of them provide Python scripts for converting models. TensorFlow.js supports models trained by TensorFlow and Keras. Keras.js supports Keras models. WebDNN supports importing models from TensorFlow, Keras, Caffe and Pytorch. With the support of using pre-trained models from other DL frameworks, the development effort can be significantly reduced.

\noindent \textbf{API to save/load model.} All frameworks that support training tasks in the browser have APIs for saving models. All frameworks have APIs for loading models.

\noindent \textbf{Support for server side (Node.js).} All frameworks are supported for Node.js. Such a feature makes it possible to offload computation inside browsers onto remote servers.

\noindent \textbf{Library size.} We list the size of the library files that need to be loaded into browsers. ConvNetJS is the smallest, which is just 33KB. TensorFlow.js and brain.js have very large size of files, which are 732KB and 819KB, respectively. Small-size libraries are better for loading applications in browsers since all the files have to be downloaded on demand.

%
%

%% file: sections/performance.tex
\begin{figure*}[t!]
  \centering
  \includegraphics[width=0.98\textwidth]{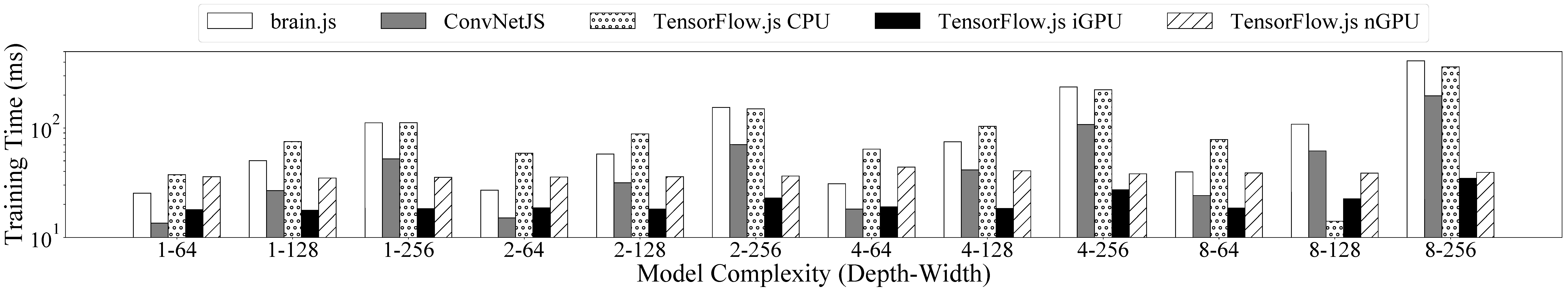}
  \vspace{-0.5em}
  \caption{Average training time (ms) on one batch under different model complexities. The y-axis is on log scale.}
  \label{fig:framework:training}
\end{figure*}

In this section, we conduct a measurement study to investigate the second research question, i.e., how well do existing frameworks perform over different DL tasks? We study the influence of model complexity and backend processor (CPU or GPU) on the performance when the browser runs training and inference tasks.

\subsection{Experiment Setup}
\noindent \textbf{DL model.}
As explained before, the network types supported by different frameworks are not the same. So we adopt the most basic fully connected neural network as the model in the experiment. For the dataset to run the DL tasks, we use the classic MNIST handwritten digit recognition database~\cite{mnist}. The model to be trained has 784 input nodes and 10 output nodes. To study the influences of model complexity on the performance, we choose different configurations of the model. The parameters include 1) the number of the hidden layers (depth) of the neural network, which ranges in [1, 2, 4, 8], and 2) the number of neurons (width) in each hidden layer, which ranges in [64, 128, 256]. The range of depth and width is set based on the assumption that client-side DL models should be of small size, being able to run on the client. In the training process, the batch size is always set to 64.

\noindent \textbf{Hardware.}
In order to study the performance difference between CPU and GPU backend, we use a Hasee T97E laptop computer, which has a standalone graphics card, Nvidia 1070 Max-Q (with 8GB GPU memory). The CPU is Intel i7-8750H, which includes an Intel HD Graphics 630, enabling us to measure the performance using integrated graphics card. In the following, we use nGPU and iGPU to denote the GPU backend on the standalone Nvidia graphics card and the integrated Intel graphics card, respectively.

\noindent \textbf{Software.}
All the experiments run on the Chrome browser (version: 71.0.3578.10 dev 64-bit) on Ubuntu 18.04.01 LTS (64-bit). For the frameworks, we use their latest published version.

\noindent \textbf{Performance measurement.}
For each DL task, we implement a Web page where the configurations of DL models can be varied through the parameters in the URL. We run each DL task on the Chrome browser, and measure the time spent on finishing the task. Since each experiment usually requires running dozens of tasks under different configurations, we developed a Chrome extension to iterate all the pages and change the configuration after one task is performed. This browser extension is also responsible for monitoring the system resource usage of the Web page. At the same time, a local server records the experimental statistics uploaded by the extension.

\subsection{Training Performance}
We select four JavaScript frameworks, brain.js, ConvNetJS, synaptic, and TensorFlow.js, which support training in browsers, to compare their performance of running training tasks. All the four frameworks can train models on the CPU backend except that TensorFlow.js is also able to use the GPU backend via WebGL. We train the defined model using each framework and obtain the average time spent on training one batch. Figure~\ref{fig:framework:training} shows the results under different model complexities. Since the training time of synaptic is about tens to hundreds of times longer than that of other frameworks, we omit the result of synaptic in the figure for better presentation but the findings are similar to other frameworks.

\begin{figure*}[t!]
  \centering
    \includegraphics[width=0.98\textwidth]{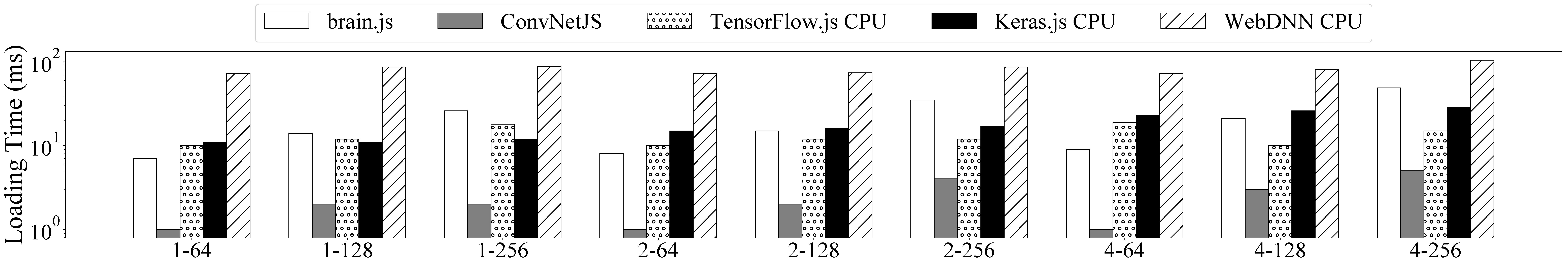}
  	\includegraphics[width=0.98\textwidth]{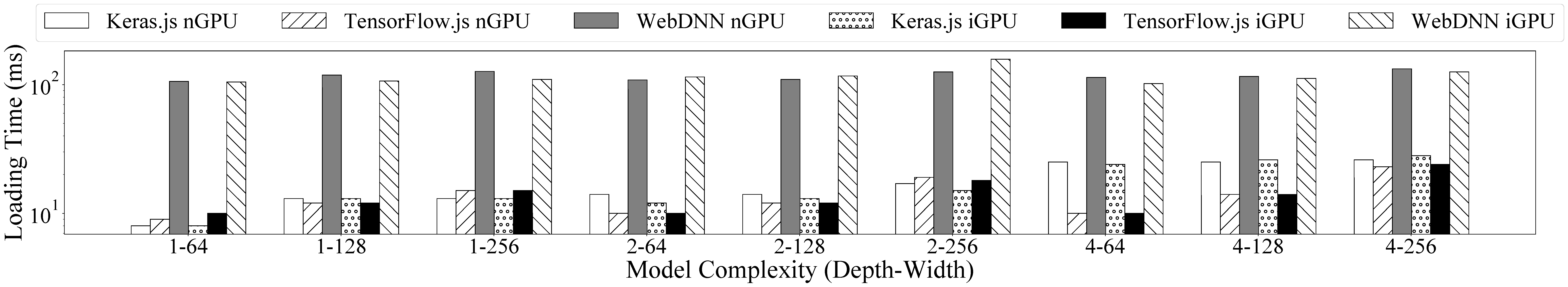}
  \vspace{-0.5em}
  \caption{Model loading time (ms) under different model complexities. The y-axis is on log scale.}
  \label{fig:framework:loading}
\end{figure*}

In general, the training time increases with the increase of the network size since more computation is needed to complete the training process for larger networks. Comparing the training time of different frameworks on the CPU backend, we can see that ConvNetJS is the fastest among all the frameworks for all network configurations. The possible reason may be that ConvNetJS is designed to be simpler, which can be reflected by its small library file size. Brain.js is closely behind, with a performance gap of about two times (2x) with ConvNetJS. Tensorflow.js has a performance gap of two to three times (2x-3x) with ConvNetJS. When comparing the training time ratio of ConvNetJS over TensorFlow.js, we find that the performance gap is gradually reduced when the depth and width increase, indicating that compared with ConvNetJS, TensorFlow.js has relatively large overhead beyond calculation. In addition, the performance gap is larger as the network width increases than as the network depth increases, implying that TensorFlow.js deals better with large-scale matrix calculation than ConvNetJS.

\noindent \textbf{GPU benefits.} The training time on the CPU backend becomes longer with the increase of network size, but the results on the GPU backend are not the same. For both the iGPU with weaker computation power and the nGPU which can satisfy larger-scale matrix calculations, the training time does not increase significantly. But in the process from (4 hidden layers, 128 neurons per layer) to (8 hidden layers, 256 neurons per layer), the training time of iGPU increases significantly. The reason may be that under the network size set in this experiment, the training process does not reach the GPU's capability bottleneck. Although the matrix computation capability of nGPU is better than that of iGPU, the training time on nGPU is even longer than iGPU. Such a result is caused by the excessive time overhead to call the WebGL for accessing GPU. The real computation time of GPU should be much shorter.

\begin{table}[t!]
\caption{CPU utilization (\%) in the training process.}\label{table:framework:CPU}
\vspace{-0.5em}
\small
\begin{tabular}{|l|r|r|r|r|}
\hline
Framework & Backend & Max & Min & Average \\ \hline
brain.js & CPU & 104.0 & 99.9 & 101.2 \\ \hline
ConvNetJS & CPU & 108.0 & 101.9 & 104.1 \\ \hline
synaptic & CPU & 113.9 & 88.7 & 102.8 \\ \hline
\multirow{3}{*}{TensorFlow.js} & CPU & 108.0 & 61.0 & 82.1 \\ \cline{2-5}
 & iGPU & 82.0 & 54.9 & 65.9 \\ \cline{2-5}
 & nGPU & 75.9 & 48.0 & 60.0 \\ \hline
\end{tabular}
\end{table}

\noindent \textbf{System resource utilization.} We show the statistics of CPU utilization of each framework during the training process in Table~\ref{table:framework:CPU}. 110\% is the upper bound of CPU utilization. The capability of multi-core processor cannot be used since the JavaScript engine is single-threaded. As a result, it can only maximize the usage of a single core. The reason why the CPU utilization is over 100\% is that other kernel and user space components occasionally run simultaneously in other threads.

On the CPU backend, TensorFlow.js sometimes cannot maximize the utilization of a single core and its average CPU utilization is only 82.1\%. Meanwhile, we can find that when running training tasks on the GPU backend, CPU is not fully utilized since most computation is on the GPU. Training on iGPU has about 5-7\% higher CPU utilization than that on nGPU.

\subsection{Inference Performance}
We select 6 JavaScript frameworks to compare their performance of running inference tasks. TensorFlow.js, Keras.js, and WebDNN support using GPU for acceleration, but brain.js, ConvNetJS, and synaptic support using only CPU for inference. In terms of model usage, brain.js, ConvNetJS, synaptic and TensorFlow.js support saving their own trained models, while Keras.js and WebDNN only support importing pre-trained models from other deep learning frameworks. Therefore, for brain.js, ConvNetJS, synaptic and TensorFlow.js, we use the models saved by the frameworks themselves. For Keras.js and WebDNN, we use the models trained by Keras and then convert the models to the corresponding format. Theoretically, the parameter values of the trained DL models should be different, but the absolute value does not affect the inference time. So we just assign the same parameter values to all the models of different frameworks.

The inference task involves loading a pre-trained model and then given a sample input, the model outputs the result. In addition, on the GPU backend, there is a warmup process where the first sample for inference is usually used to activate the GPU processor. Therefore, we break down the inference process into three phases: model loading, warming up, and inference, and study the fine-grained performance. Due to the space limitation, we omit the results where the model depth is 8 in the following analysis because the trend is similar as the depth increases. Besides, since the model loading time and inference time of synaptic are still much longer than those of other frameworks, we do not depict the results of synaptic in the figures for better presentation.

\noindent \textbf{Model file size.} We first investigate the size of the model file used by different frameworks. As models for inference usually should be downloaded from the remote server, smaller size of model files means shorter downloading time. Table~\ref{table:framework:modelsize} shows the size of model files that are used in all inference experiments. ConvNetJS and brain.js use similar JSON encoding, so the size of their model files are nearly the same. The model file of synaptic uses JSON encoding as well but its size is the largest among all the frameworks. As the model files used by TensorFlow.js, Keras.js and WebDNN are all converted from Keras models, their model files are of the same size. So we just show TensorFlow.js in the table.  Since the model converted from Keras is compressed and saved as a binary file, the size can be greatly reduced, just about 1/7 of the model file in JSON.

\begin{table}[t!]
\caption{Size of model files (MB).}\label{table:framework:modelsize}
\vspace{-0.5em}
\small
\begin{tabular}{|r|r|r|r|r|r|}
\hline
Depth      & Width & brain.js & ConvNetJS & synaptic & TensorFlow.js \\ \hline
\multirow{3}{*}{1} & 64      & 1.4      & 1.3       & 3.4      & 0.2                           \\ \cline{2-6}
                   & 128     & 2.7      & 2.7       & 6.7      & 0.4                           \\ \cline{2-6}
                   & 256     & 5.5      & 5.4       & 13.3     & 0.8                           \\ \hline
\multirow{3}{*}{2} & 64      & 1.5      & 1.5       & 3.7      & 0.2                           \\ \cline{2-6}
                   & 128     & 3.2      & 3.1       & 7.8      & 0.5                           \\ \cline{2-6}
                   & 256     & 7.2      & 7.1       & 17.7     & 1.1                           \\ \hline
\multirow{3}{*}{4} & 64      & 1.7      & 1.7       & 4.2      & 0.3                           \\ \cline{2-6}
                   & 128     & 4.0      & 4.0       & 10.1     & 0.6                           \\ \cline{2-6}
                   & 256     & 10.7     & 10.5      & 26.5     & 1.6                           \\ \hline
\end{tabular}
\end{table}

\begin{figure*}[t!]
  \centering
  \includegraphics[width=0.98\textwidth]{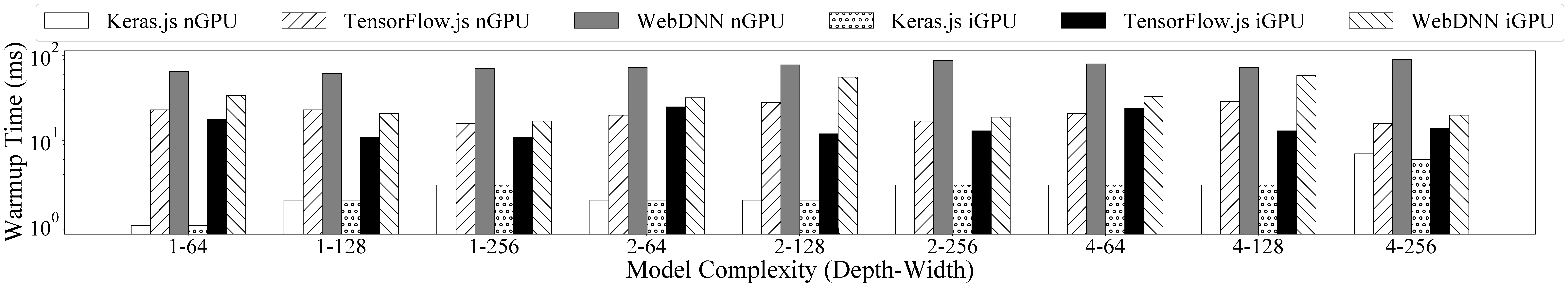}
  \vspace{-0.5em}
  \caption{Model warmup time (ms) on GPU under different model complexities. The y-axis is on log scale.}
  \label{fig:framework:warmup}
\end{figure*}

\begin{figure*}[t!]
  \centering
    \includegraphics[width=0.98\textwidth]{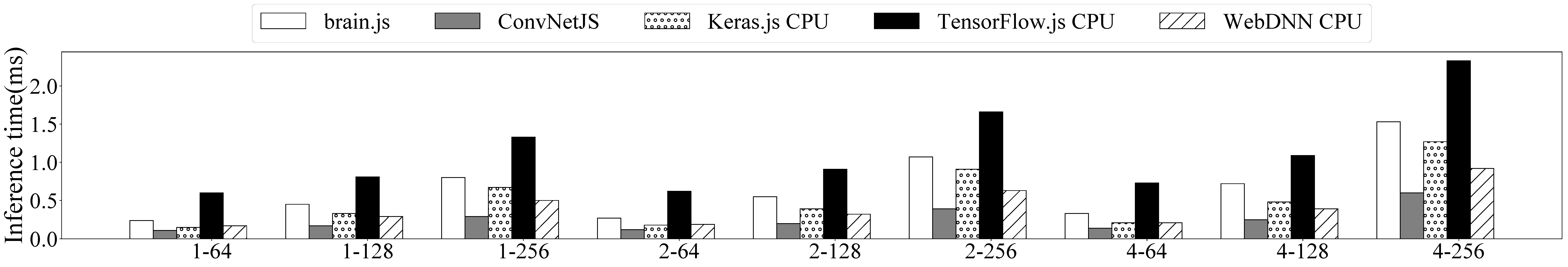}
  	\includegraphics[width=0.98\textwidth]{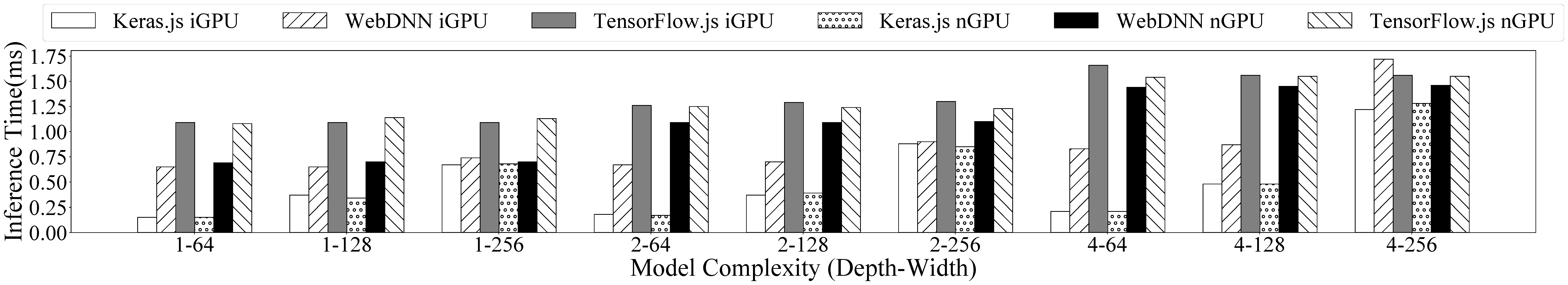}
  \vspace{-0.5em}
  \caption{Average inference time (ms) on one sample under different model complexities.}
  \label{fig:framework:inference}
\end{figure*}

\noindent \textbf{Model loading time.} We then compare the time spent on loading the model of different frameworks, as shown in Figure~\ref{fig:framework:loading}. For the CPU backend, the loading time of different models of the same framework is proportional to the size of the model files described in Table~\ref{table:framework:modelsize}. However, the model loading time of different frameworks is significantly different. ConvNetJS is the fastest. Model loading time of brain.js, TensorFlow.js and Keras.js are consistent in terms of magnitude. Interestingly, the increase of loading time of ConvNetJS, brain.js and synaptic is particularly noticeable when the width increases. The result is caused by their choice of using JSON to encode models. The model loading time of synaptic is slowest among all the frameworks, which are more than 100x to 1000x longer than ConvNetJS. The model loading time of TensorFlow.js is almost unchanged regardless of the model size.

The loading time on the GPU backend does not change much under different model complexities. However, the difference is still significant between different frameworks. TensorFlow.js is the fastest. Compared with loading models on the CPU backend, Keras.js speeds up loading large models, but the loading time of WebDNN is longer. In addition, it can be seen that there is no difference in the model loading time between iGPU and nGPU.


\noindent \textbf{Warmup time.} Next, we examine the difference of warmup time on the GPU backend. As shown in Figure~\ref{fig:framework:warmup}, Keras.js is still far ahead, and can complete the warmup in 3ms on all tasks. Tensorflow.js is the second, and WebDNN is the worst. On the whole, the warmup time on iGPU backend is shorter than that on nGPU.


\noindent \textbf{Inference time.} Figure~\ref{fig:framework:inference} shows the average time of doing inference on one sample. Almost all the inference tasks can finish within 1.5ms (except synaptic, of which the shortest is 6.68ms). In the range of the model sizes we set, the powerful computation capability of GPU does not make a difference. Among all the model sizes, ConvNetJS occupies all the first place, followed by WebDNN on the CPU backend. The inference time of WebDNN on the GPU backend is longer than the inference time on the CPU backend. As for TensorFlow.js, running on the CPU backend is faster for inference on smaller models, while the GPU backend is faster for inference on larger models. Inference times of Keras.js on the CPU and GPU backend are basically the same.

We can observe that for all the frameworks on the CPU backend, the inference time increases when the model becomes complex. In particular, when the width increases, the time increases sharply (about two times as the model width doubles). Similar to the training tasks, such a result also reflects that these frameworks do not optimize the large-scale matrix operations in the process of forward propagation on the CPU backend. TensorFlow.js and WebDNN on the GPU backend do not exhibit this problem, but Keras.js on the GPU still suffers from this problem.

\subsection{Takeaway}
Based on the above results, we can see that in small-scale fully-connected neural network which the browser is capable of, ConvNetJS performs the best for both training and inference. However, since ConvNetJS is no longer maintained and has fewer functionalities, developers may need to choose some alternatives.

Tensorflow.js is the only framework that can take advantage of GPU to accelerate training processes. It is feature-rich and has comparable performance with ConvNetJS. So TensorFlow.js is a good choice for both training and inference. We do not recommend using GPU as the backend on small models because the advantage of GPU's computation power is not fully exploited.

Finally, we are interested in why ConvNetJS has the best performance for all the tasks among these frameworks. Given the same model of which the process logic is the same, the performance difference is likely to be accounted by the different implementation details. To this end, we compare the function call stack of ConvNetJS with that of TensorFlow.js when doing the same training task. It is surprising to find that the depth of the call stack of ConvNetJS is only 3 while TensorFlow.js is 48! Such a result suggests that one possible reason for the performance difference among different frameworks is the deep call stack that costs a lot of computation resources.

%% file: sections/comparison.tex
In this section, we study the third research question, i.e., how big is the performance gap between running DL in the browser and on the native platform? To this end, we compare the performance of TensorFlow.js and the native TensorFlow in Python, both of which are released and maintained by Google and have similar APIs, making the comparison fair enough.

We study the performance gap from two aspects. On one hand, we leverage well-known pre-trained models to compare the performance when running inference tasks on TensorFlow.js and native TensorFlow. On the other hand, we use decision tree analysis to distinguish the factors contributing to the performance gap. We use the same laptop as the one used in the experiments of the last section. We install the latest TensorFlow in Python (version 1.11.0) on the laptop.

\subsection{Inference Based on Pre-Trained Models}
We use the pre-trained models officially provided by the Keras to measure the performance of TensorFlow.js and native TensorFlow when doing inference tasks on these classical models.

\subsubsection{Limitations of TensorFlow.js and browser constraints}
Keras officially provides 11 pre-trained models. Although these models can work using native TensorFlow, we encountered a series of errors when we run them using TensorFlow.js in the browser. These errors imply the limitations of TensorFlow.js itself as well as constraints imposed by the browser.

For the model of NasNet Large, the browser throws out the error message ``truncatedNormal is not a valid Distribution''. For the model of ResNet V3, the browser throws out the error message ``Unknown layer: Lambda''. The reason for these two errors is that TensorFlow.js is still under development and so far has offered only a limited number of support for the converted model. Many user-defined operations are not supported by TensorFlow.js, e.g., models with control flow operations in RNNs are not yet supported.

When we try to use VGG16 or VGG19, the browser throws out the error message ``GL OUT OF MEMORY'', meaning that the GPU memory is overfilled. The reason is that the VGG16 model applies for more than 1GB GPU memory. However, it should not be an issue since the GPU memory of our experiment laptop is 8GB. As a result, such an error is due to the browser constraints.

\begin{table}
\caption{Selected Keras pre-trained models.}\label{table:pretrained}
\vspace{-0.5em}
\small
\begin{tabular}{|c|r|r|r|}
\hline
Model Name  & \begin{tabular}[c]{@{}l@{}}Pre-trained \\ Model Size\end{tabular} & \begin{tabular}[c]{@{}l@{}}Trainable \\ Parameters \end{tabular} & \begin{tabular}[c]{@{}l@{}}Computation\\ (FLOPs) \end{tabular}  \\
\hline
MobileNetV2 & 14MB                  & 3.5M                 & 7.2M                \\
\hline
DenseNet121 & 33MB                  & 8.0M                 & 16.3M               \\
\hline
Xception    & 88MB                  & 22.9M                & 46.0M               \\
\hline
InceptionV3 & 92MB                  & 23.8M                & 47.8M               \\
\hline
ResNet50    & 99MB                  & 25.6M                & 51.4M               \\
\hline
\end{tabular}
\end{table}

After trying all the models, we finally have 5 models that can be correctly converted and run on the browser. The information of these models are listed in Table~\ref{table:pretrained}. The number of trainable parameters is obtained by the build-in \texttt{summary()} method of tensorflow.keras, and the computation complexity (Floating Operations) are obtained by \texttt{tensorflow.propfiler.profile()} method.

\subsubsection{Results}
\begin{figure}[t!]
	\includegraphics[width=0.45\textwidth]{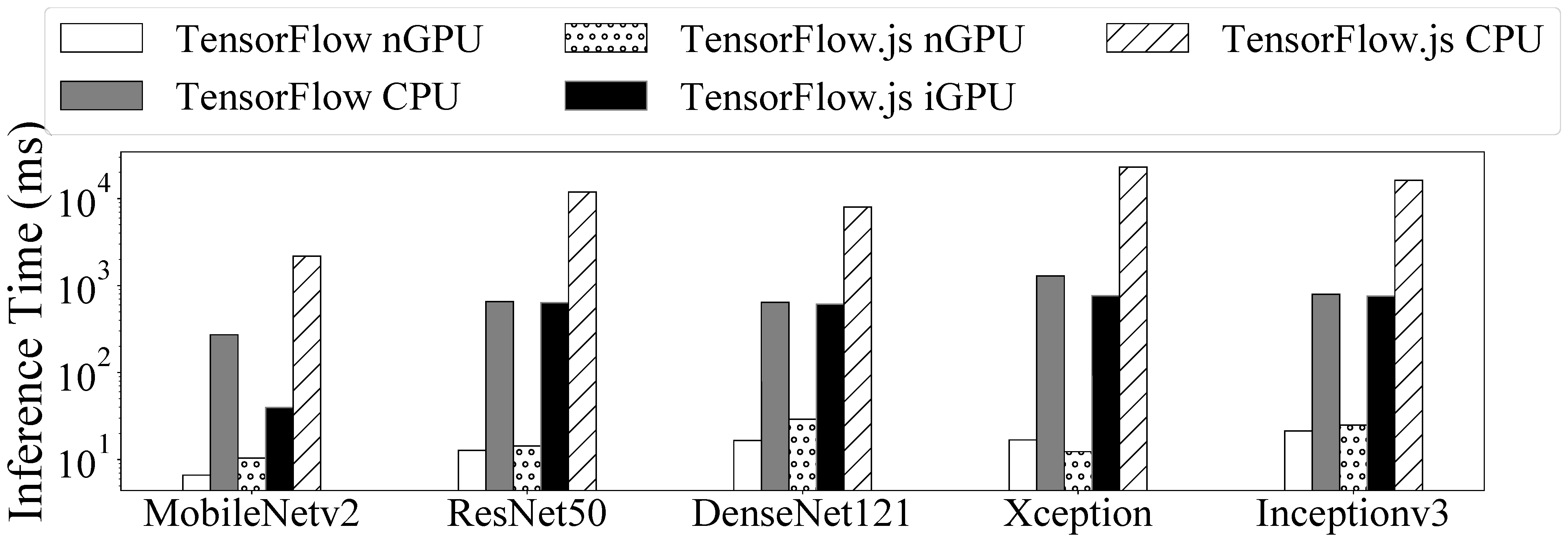}
    \vspace{-0.5em}
	\caption{Inference time on pre-trained Keras models. The y-axis is on log scale.}\label{figure:pretrained:time}
\end{figure}
Figure~\ref{figure:pretrained:time} shows the inference time for each model. It can be seen that the inference time of TensorFlow.js on nGPU is comparable (1x-2x slower) to native TensorFlow's. The most encouraging result is that the performance of TensorFlow.js on the iGPU backend is better than that of native TensorFlow on the CPU backend. This result is not surprising considering the computation capability of integrated graphics card and CPU. However, since traditional native DL frameworks do not support integrated graphics card for acceleration, DL tasks can benefit a lot from browsers in such a case with the help of integrated graphics card that is common on current devices.

Under the real-time requirement of client-side DL, if users want to achieve the experience of 10FPS (frame per second), they need to consider using a more powerful standalone graphics card. The Mobile Net model accelerated by iGPU can also meet the requirement. If the requirement is 1FPS, iGPU is also fully capable. But if only CPU can be used, then these common models are too heavy to run in browsers.

\subsection{Decision Tree Analysis}
In order to deeply reveal how different factors of DL tasks influence the performance gap between DL in browsers and on native frameworks, we build a predictive model based on decision tree analysis to study the factor importance.

\begin{table}[t!]
\caption{Contributing factors to the performance gap.}
\vspace{-0.5em}
\small
\label{table:params_config}
\begin{tabular}{|c|l|l|}
\hline
Network Type & Factor & Range \\ \hline
\multirow{4}{*}{DNN} & Backend &  CPU, GPU  \\ \cline{2-3}
& Task Type &  training, inference  \\ \cline{2-3}
& Depth & 1, 2, 4, 8, 16  \\ \cline{2-3}
& Width & 64, 128, 256, 512 \\ \hline
\multirow{4}{*}{CNN} & Backend &  CPU, GPU  \\ \cline{2-3}
& Task Type &  training, inference  \\ \cline{2-3}
& Depth & 6, 9, 15, 27  \\ \cline{2-3}
& Width & 200, 400, 800 \\ \hline
\multirow{4}{*}{RNN} & Backend &  CPU, GPU  \\ \cline{2-3}
& Task Type &  training, inference  \\ \cline{2-3}
& Depth & 1, 2, 3  \\ \cline{2-3}
& Width & 4, 8, 16, 32, 64, 256 \\ \hline
\end{tabular}
\end{table}

\begin{figure*}
  \centering
    \subfigure[DNN]{
    \label{fig:tree_dnn}
    \includegraphics[width=0.32\textwidth]{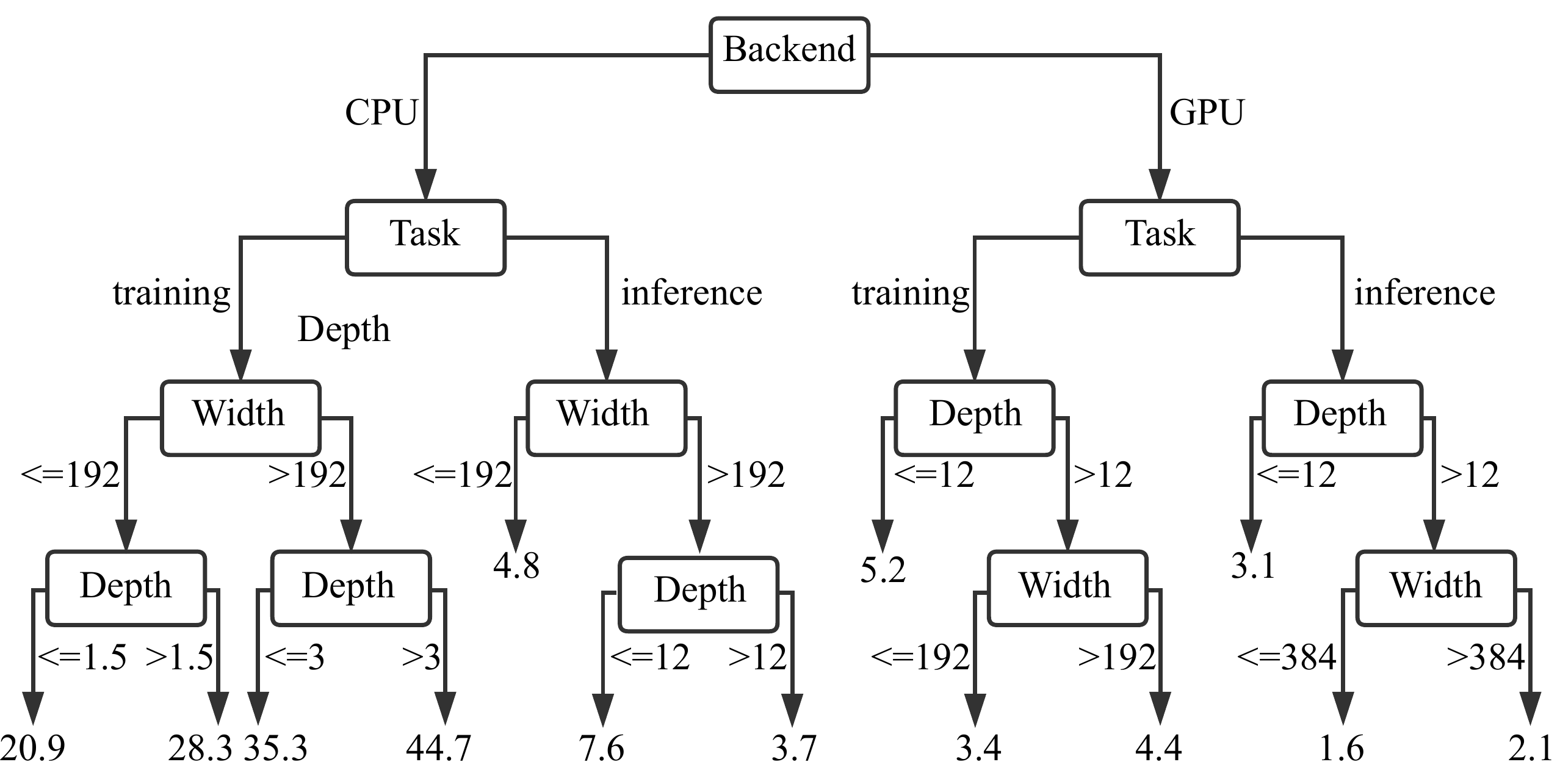}}
    \subfigure[CNN]{
    \label{fig:tree_cnn}
    \includegraphics[width=0.33\textwidth]{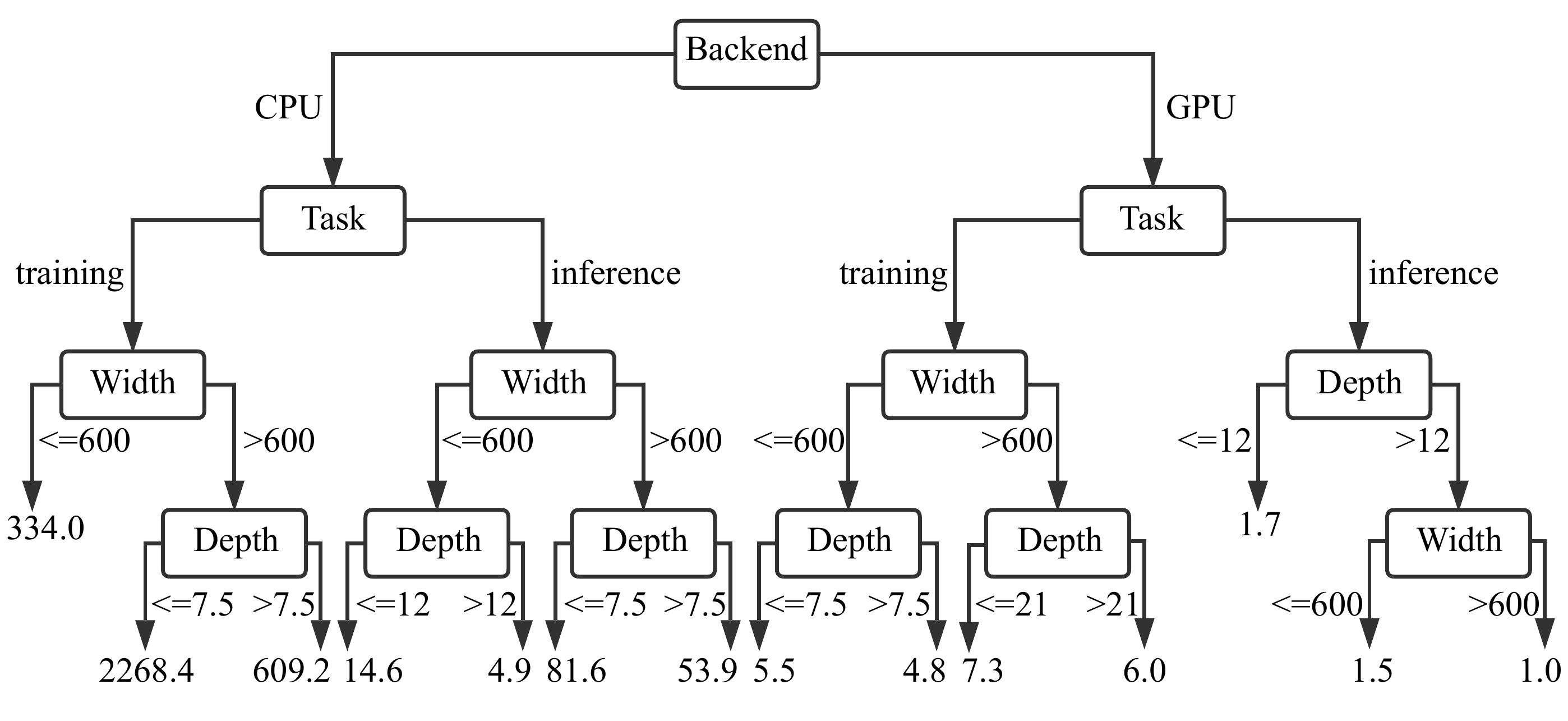}}
    \subfigure[RNN]{
    \label{fig:tree_rnn}
    \includegraphics[width=0.32\textwidth]{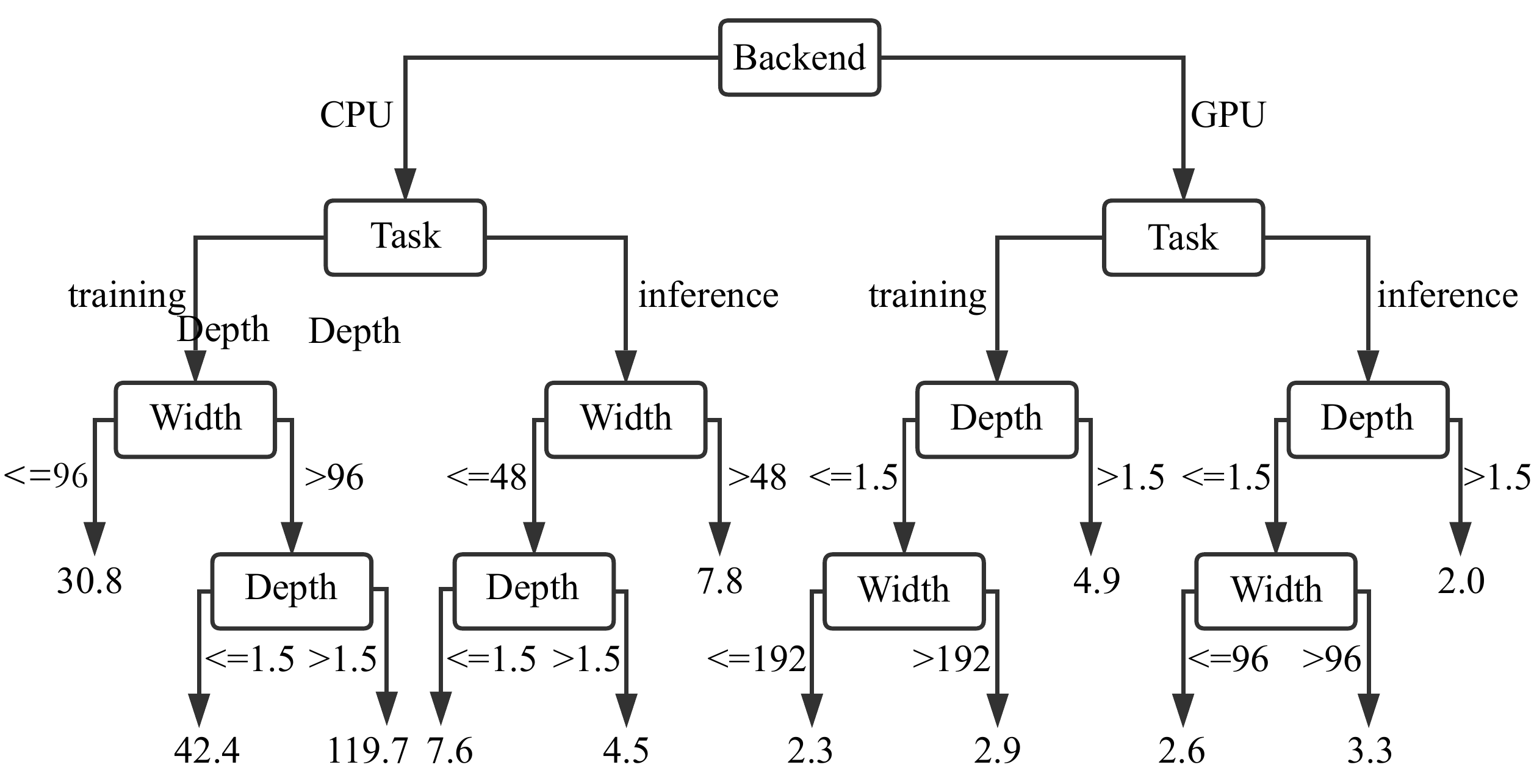}}
    \vspace{-1em}
  \caption{Decision tree to analyze the time ratio of TensorFlow.js over native TensorFlow on DNN, CNN, and RNN Models.}
  \label{figure:decisiontree}
\end{figure*}

\subsubsection{Experiment Setup}
We consider 4 factors that influence the performance gap between DL in browsers and on native platforms as shown in Table \ref{table:params_config}, including backend (CPU or GPU), task type (training or inference), as well as depth and width that represent the model complexity. In the DNN and RNN models, width refers to the number of neurons of each layer. In the CNN models, width refers to the number of kernels used in the convolution layer. For each of DNN, CNN and RNN, we choose one model from the Tensorflow.js official examples. The DNN and CNN models are used to recognize handwritten digits on the MNIST dataset, and the RNN model is to perform text generation from Nietzsche's writings. The range of depth and width is selected according to the values set in Tensorflow.js official examples.

In our experiment, we build and run the DNN, CNN and RNN models under different configurations using TensorFlow.js and native TensorFlow, respectively. Each configuration is a combination of values for the factors above. We measure the execution time of each configuration as the average time per batch for training tasks and average time per sample for inference tasks on two platforms. We use the ratio of the execution time on TensorFlow.js over that on native TensorFlow to quantify the performance gap.

\subsubsection{Methodology}
We run the decision tree algorithm with sklearn~\cite{sklearn} to predict the ratio of execution time between TensorFlow.js and native TensorFlow. The decision tree depicts the relative importance of contributing factors. Intuitively, factors close to the root of the decision tree affect the time ratio more than those near the leaves. This is because the decision tree chooses to do the splitting of the nodes according to the Entropy-Information Gain criterion. In other words, the decision tree places the important factors near the root to gain the best prediction of splits.

Based on the results, we first produce a fully grown and unpruned decision tree for all the factors. In this way, each leaf contains only one configuration. Then we set the depth of the tree to the number of factors, in order to prevent using a factor several times on one path. Figure \ref{figure:decisiontree} shows the decision trees for DNN, CNN, and RNN.


\subsubsection{Results}

The execution time of TensorFlow.js is longer than native TensorFlow in almost every configuration.

Backend is the most important factor contributing to the performance gap. The ratio of execution time on the CPU backend is much higher than that on the GPU backend. For example, the ratio decreases from 44.7 to 4.4 for training tasks when the DNN model with depth over 3 and width over 192 runs on the GPU backend instead of on the CPU backend. The extreme case happens on the CNN. On the CPU backend, there is a wide range of the ratio from below 5 to over 2200 (when depth is less than 7.5 and width is over 600). However, when doing inference task on the GPU backend with depth over 12 and width over 600, TensorFlow.js performs as fast as native TensorFlow. This is because CNN makes use of the powerful computation capability of GPU when the model is large enough, yet not exceeding the upper bound of the browser memory.

The second most important factor is task type for all the three models. Performing training tasks exhibits a higher ratio, while the performance gap on inference tasks is small. For example, for the DNN model on the CPU backend, training tasks of TensorFlow.js perform 33.9 times slower than native TensorFlow on average, and inference tasks of TensorFlow.js perform 5.8 times slower than native TensorFlow on average.

The decision trees of DNN and RNN both suggest that the importance of depth and width depends on which backend the task is taken on. On the CPU backend, the importance of width outweighs that of depth, while depth plays a more important role on the GPU backend. However, in the case of CNN, width plays a more important role to the performance gap than depth for training tasks.

%% file: sections/implication.tex
Table~\ref{table:implication} summarizes the findings and implications of our study. Specifically, we draw implications for three stakeholders of DL in browsers: application developers, DL-framework vendors, and browser vendors. For application developers, we give recommendations on how to choose frameworks for DL in browsers, how to optimize the model, as well as how to select the backend. For DL-framework vendors, we present some advice on encoding of model files and optimizing the call stack. For browser vendors, we suggest on the utilization of system resources.

\begin{table*}[t!]
\caption{Major findings and implications of DL in browsers.} \label{table:implication}
\vspace{-0.5em}
\footnotesize
\begin{tabular}{|p{1em}|p{7em}|p{30em}|p{22em}|p{5em}|}
\hline
\textbf{No.} & \textbf{Name}               & \textbf{Finding}                                                                                                                                                                                                                                                                                  & \textbf{Implication}                                                                                                                                                             & \textbf{Stakeholder}  \\ \hline
1            & Specific DL Tasks Support          & Frameworks supporting DL in browsers are emerging and being actively maintained. Most of them are not for general purpose and support only a specific subset of DL tasks. & It is better for developers to use general-purpose DL frameworks like TensorFlow.js to implement their DL-powered Web applications.                                                             & Application Developer \\ \hline
2            & Model Complexity             & The width of DL models dominates the performance variation of both training and inference tasks considering the complexity of DL models.                                                                                                                                                          & Developers should pay attention to the width of their models, and balance the width and required performance if possible.                                                                                  & Application Developer \\ \hline

3            & Model Loading               & For inference tasks, loading and warming up the DL model accounts for much longer time than running the inference task itself. The warmup time on the integrated graphics card is generally shorter than that on the standalone graphics card.                                                                              & Developers should pre-load and warm up the model before using it for inference.                                                                                                  & Application Developer \\ \hline
4            & Benefits from GPU                & For popular pre-trained models like MobileNet and Inception, TensorFlow.js has comparable performance with native TensorFlow when running inference on the standalone graphics card.                                                                                                                     & It is possible to develop Web applications rather than native applications for these tasks.                                                                                    & Application Developer \\
\hline
5            & Benefits from Integared Graphics Card               & TensorFlow.js running on the integrated graphics card works better than native TensorFlow running on CPU backend.                                                                                                                                                                                     & For devices without standalone GPUs, developers can use the browser for DL tasks, leveraging integrated graphics card for acceleration.                                                                                      & Application Developer \\
\hline
6            & Model File Encoding and Size             & Model file encoded in JSON is much bigger (7x) in size than that encoded in binary, and significantly increases the model loading time.                                                                                                                                  & It is better to encode DL models in binary files.                                                                     & DL-Framework Vendor   \\
\hline
7            & Framework Call Stack        & The call stack of TensorFlow.js is much deeper than that of ConvNetJS, pulling down the performance.                                                                                                                                                                               & Framework vendors could leverage compiler optimization techniques to reduce the call stack when the DL models are used in the production environment.  & DL-Framework Vendor   \\
\hline
8            & System Resource Utilization & The capability of multi-core CPU cannot be utilized when running DL tasks on the CPU backend in browsers since the JavaScript program is single-threaded. GPU memory usage is limited in 1GB, failing to load and run larger models.                                                                  & JavaScript engine should take into account the support of multi-process or scheduling among multi cores for better performance of DL tasks in browsers. The GPU memory should be configurable for DL tasks. & Browser Vendor        \\
\hline

\end{tabular}
\end{table*}

%% file: sections/related.tex
To the best of our knowledge, this paper is the first study to characterize the DL in browsers. So we survey related work on general client-side DL and performance measurement of DL systems.




\subsection{Client-side Deep Learning}
With the emphasis on privacy, personalization and timely response, it is a trend to conduct DL directly on the clients where the data is generated, especially on mobile devices. Lane et al.~\cite{lane2015can} studied the feasibility of using DL for typical mobile sensing tasks, such as activity recognition. Yao et al.~\cite{Yao:2017:DUD:3038912.3052577} proposed DeepSense, a unified DL framework for processing time-series mobile sensing data. Despite of the increasing computation power of mobile devices, typical DL tasks are still of heavy workload for these resource-constraint devices. Several optimization methods were proposed to improve the performance of client-side DL. One line of work focuses on the DL models. Han et al.~\cite{Han:ICLR2016} proposed deep compression to compress the DNN through a three-stage method: pruning, trained quantization and Huffman coding, which showed a considerable reduction in terms of the storage requirements of DNNs. The other line of work leverages the cloud and edge environment to offload the computation-intensive tasks to powerful computation nodes~\cite{liu2017appbooster}. Kang et al.~\cite{Kang:ASPLOS2017} proposed Neurosurgeon, a lightweight scheduler to automatically partition DNN computation between mobile devices and data centers at the granularity of neural network layers. Wang et al.~\cite{DBLP:conf/kdd/WangZBZCY18} designed Arden, a cloud-based deep learning framework for mobile devices. The framework partitions the DNN and offloads the resource-hungry training and complex inferences tasks to the cloud.

Anther usage scenario of client-side DL is to support the distributed deep learning. Teerapittayanon et al.~\cite{teerapittayanon2017distributed} proposed distributed deep neural networks (DDNNs) over distributed computing hierarchies, consisting of the cloud, the edge (fog) and end devices. Ichinose et al.~\cite{DBLP:journals/ieicet/IchinoseTNO18} proposed a pipelined method for distributed DL processing between mobile devices and the cloud to reduce the amount of data sent to the cloud and protect the privacy of users. Meeds et al.~\cite{10.7717/peerj-cs.11} designed MLitB, a prototype DL framework capable of performing large-scale distributed computing with heterogeneous classes of devices using Web browsers.

\subsection{Performance Measurement of Deep Learning}
In recent years, researchers have conducted studies to measure the performance for various kinds of deep learning tasks. Liu et al.~\cite{Liu2018Performance} evaluated the performance of leading DL methods for object detection. Guignard et al.~\cite{DBLP:conf/hicss/GuignardSBWV18} presented detailed characterization results of a set of archetypal state-of-the-art DL workloads to identify the performance bottlenecks and to guide the design of prospective acceleration platforms in a more effective manner. Shi et al.~\cite{DBLP:conf/dasc/ShiWC18} evaluated the performance of four state-of-the-art distributed DL frameworks over different GPU hardware environments. They built performance models of standard processes in training DNNs with SGD, and then benchmark the performance of the frameworks with three neural networks (i.e., AlexNet, GoogleNet and ResNet-50). As for DL on mobile devices, Ignatov et al.~\cite{ignatov2018ai} studied state-of-the-art DL in the Android ecosystem and described available frameworks, programming models and the limitations of running AI on smartphones.

Although many JavaScript-based frameworks have been published to support DL in browsers, there is no comprehensive study to understand their characteristics and performance. Some researchers focus on the possibility of supporting DL in browsers by measuring the low-level browser capabilities. Malle et al.~\cite{malle2018need} presented a comparison study between native code and different browser-based implementations: JavaScript, ASM.js as well as WebAssembly on a representative mix of algorithms. However, these algorithms are not DL tasks. Their goal is just to show that the browsers performance is now comparable to and even exceeds native binary performance.

%% file: sections/conclusion.tex
This paper made the first study on understanding the feasibility and performance of deep learning in Web browsers. We chose 7 recently emerging JavaScript-based DL frameworks and comprehensively revealed which type of DL tasks have been supported. We measured the performance of different frameworks when doing different DL tasks in browsers, and compared with the native DL framework to investigate the performance gap. Although the in-browser DL is still at the early stage, some interesting findings, e.g., the comparable performance of JavaScript frameworks to that of native ones on some types of DL tasks and the benefits gained from the integrated graphics card, can be useful and help guide the DL-powered Web applications. Additionally, we have also found that there are some potential space of improvement for currently in-browser DL frameworks, and plan to realize some practical solutions. We believe that our work can shed a light on the future of Web applications in the AI era.